\documentstyle[12pt]{article}
\begin{document}

\baselineskip 0.9cm

\begin{center}
{\LARGE  Local and global structure of a thick domain wall space-time}
\end{center}

\centerline {\it Anzhong Wang\footnote{e-mail address: wang@on.br}}

\begin{center}
Department of Astrophysics, Observatorio Nacional, Rua General
Jos\'e Cristino $77, 20921-400$ Rio de Janeiro, RJ, Brazil
\end{center}

\centerline {\it Patricio S. Letelier\footnote{e-mail address:
letelier@ime.unicamp.br} }
\begin{center}
Department of Applied Mathematics-IMECC, Universidade Estadual de Campinas,
$13081-970$ Campinas, SP,  Brazil
\end{center}

\vspace{1.cm}

\baselineskip 0.6cm

The local and global properties of the Goetz thick plane domain wall
space-time are studied. It is found that when the surface energy of the wall is
greater than a critical value $\sigma_{c}$, the space-time will be closed
by intermediate singularities at a finite proper distance. A model is
presented in which these  singularities will give rise to scalar ones
when interacting with  null fluids. The maximum
extension of the space-time of the wall whose surface energy is less than
$\sigma_{c}$ is presented. It is shown that for certain choice
of the free parameter the space-time has a black hole structure but with
plane symmetry.

\vspace{1.cm}

\noindent PACS numbers: 04.20.Dw, 04.30.Nk, 97.60.Sm.

\newpage

\baselineskip 0.9cm

The study of topological domain walls is crucial for understanding
the inflationary Universe scenario. According to the theory [1], the
Universe undertakes an exponential expansion in the early stages of
its evolution, triggered by some
phase transitions associated with spontaneous symmetry breaking of
the Higgs field of some Grand Unified Theory. It was found
that such a kind of expansions is needed in order to solve some
long-standing puzzles in Cosmology, such as, the horizon, flatness,
and monopole. The evolution of the Universe from the exponential
expansion to its present Friedmann-Robertson-Walker form is completed
by the spontaneous nucleation of bubbles of true vacuum. Guth's original
idea has experienced several complementary  modifications [2]. It is the
general belief that inflation will finally solve the above mentioned
problems  of the standard big-bang Cosmology.

The work of  Hill, Schramm, and Fry (HSF) [3], renewed the interest in
the cosmological significance of domain walls.
In the HSF model, the phase transition happens after the time of
recombination of matter and radiation. So, domain walls produced during
this phase transition are very light and thick, and are assumed to
provide the gravitational field necessary for the clustering of dark
matter and baryons after the recombination. Those ``soft" walls have
promising implications to the Universe [4]. Following this line, an
interesting solution of thick plane domain wall to the Einstein field
equations has been found recently by Goetz [5] and rederived by
Mukherjee [6].

Motivated by the inflationary Universe scenario and the interest of the
walls on their own right, in this paper we shall present a detailed
study of the Goetz solution and pay main attention
on its global properties. It is found that for some choice of the free
parameter $q$ appearing in the solution, the space-time has a black hole
structure but with plane symmetry. It has two
asymptotically flat regions separated from the catastrophic ones
by event horizons. In each of the two asymptotically flat
regions there is a wall. However, a more careful study shows
that a Minkowski-like observer will find that the space-time actually has
spherical symmetry, and that the walls are bubbles which are collapsing
initially and expanding later with a constant acceleration. In this respect,
we can see that the thick domain wall shares the same property as those
with zero-thickness [7, 8]. Examples of single solutions with  several
different physical interpretations are not rare in General Relativity [9].
It should be noted that domain walls with a  causal lattice structure
similar to the Reissner-Nordstrom and Kerr black holes have been found in
Supergravity [10], and some interesting features have been obtained [10, 11].

The Goetz solution can be written as [5, 6]
\begin{equation}
ds^2 = e^{-\Omega} (dt^2 - dz^2) - e^{-h}( dx^2 + dy^{2} ),
\end{equation}
with
\begin{equation}
\Omega = 2q\;ln(cosh(pz)), \;\;\;\; h = 2q\;ln(cosh(pz)) - 2kt,
\end{equation}
where $p \equiv k/q,$ and $k$ and $q$ are arbitrary constants subject to the
conditions $k > 0,$ and $0 < q <1.$ The coordinates take the range
$- \infty < t, z, x, y < + \infty$. From Eq.(1) we see that
Goetz wall is plane symmetric  with the Killing vectors,
 $\partial_x, \partial_y,$ and $y\partial_x - x\partial_y$.
The corresponding energy-momentum tensor (EMT) is given by [12]
$T_{\mu \nu} = \rho (g_{\mu \nu}
 + \xi_{\mu} \xi_{\nu}) + \nu \xi_{\mu} \xi_{\nu},$ where $\rho$ denotes
the energy
density of the wall, $\nu$ the pressure in the direction perpendicular to
the wall, and are given by
\begin{equation}
\rho = -\left(\frac{q + 2}{3q}\right) \nu = k^{2}\left(\frac{q + 2} {q} \right)
\{cosh(pz)\}^{-2(1-q)}.
\end{equation}
The unity vector $\xi_{\mu}$ is the normal to the wall and given by
$\xi_{\mu} = e^{-\Omega/2} \delta^{z}_{\mu}$. As shown in [5, 6], the
above solution corresponds to a Higgs scalar field with a kinklike
shape $\phi \;( = Arctan[sinh(pz)]), $  self-interacting through
the potential $V(\phi) = \{cos^{2}\phi\}^{(1-q)}$.
An interesting feature of this solution is the existence of coordinate
singularities in each of the three spatial directions. The ones in the $x$
and $y$ directions are obvious, since the hypersurfaces $z = Const.$ are
the $(2+1)-$dimensional de Sitter spaces. Thus, these coordinate singularities
are the usual de Sitter horizons. The coordinate singularities in
the $z$ direction
can be seen by computing the proper distance between $z = 0$, the center
of the wall, and $|z| = \infty$, which is found finite [5].
Consequently, the hypersurfaces $|z| = \infty$ represent coordinate
singularities,
too. Therefore, in general an extension in each of these three directions
is needed. The extension in the $x$ and $y$ directions is simple and
similar to its four-dimensional analog of the de Sitter space-time given
in [13]. Thus, in the following we shall restrict ourselves only to the
extension in the $z$ direction.

  It will be useful to consider first the hypersurfaces $|z| = \infty$ in more
 details, specially  the timelike geodesics perpendicular to the wall.
 From the first
integral, it is found [5, 6] that the time-like geodesic equations yield
$dt/d\tau = E cosh^{2q}(pz), dz/d\tau = \pm
cosh^{q}(pz) \{E^{2} cosh^{2q}(pz) - 1\}^{1/2}, dx/d\tau = 0,$ and
$dy/d\tau = 0,$ where $E$ is the energy of the test particle, and $\tau$
the proper time. Perpendicular to the timelike vector $\lambda^{\mu}_{(0)}
(\equiv
dx^{\mu}/d\tau),$ we have  other three linearly independent
spacelike vectors $\lambda^{\mu}_{(a)} ( a = 1, 2, 3)$ defined by
$\lambda^{\mu}_{(1)} =
(dz/d\tau)\delta^{\mu}_{t} + (dt/d\tau)\delta^{\mu}_{z},
\lambda^{\mu}_{(2)} = e^{h/2} \delta^{\mu}_{x},$ and $\lambda^{\mu}_{(3)}
 = e^{h/2} \delta^{\mu}_{y},$ where $h$ is given by Eq.(2). One can show
that such defined four unity vectors form an orthogonal tetrad and have the
properties $\lambda^{\mu}_{(0)};_{\nu} \lambda^{\nu}_{(0)} = 0 =
\lambda^{\mu}_{(a)};_{\nu} \lambda^{\nu}_{(0)}.$
Thus, the three vectors $\lambda^{\mu}_{(a)}$ are parallel transported
along the timelike geodesics, and together with
$\lambda^{\mu}_{(0)}$ form a freely-falling frame. Computing the Riemann
tensor in this frame, we find that it has only four independent components,
one of which is given by
$$
R_{\mu \nu \sigma \delta} \lambda^{\mu}_{(0)} \lambda^{\nu}_{(2)}
\lambda^{\sigma}_{(0)} \lambda ^{\delta}_{(2)} = \frac{k^2}{q}
(cosh(pz))^{2(q-1)} \left\{ E^{2}(1-q) cosh^{2q}(pz) - 1 \right\}.
$$
Clearly, as $|z| \rightarrow \infty$, this component becomes unbounded for
$q > 1/2$. From the geodesic
equation we get that as $|z| \rightarrow \infty$ ,
$\,\,e^{2pqz}\sim (\tau_\infty -\tau)^{-1}.$  Thus the freely-falling
observer experiences a tidal force $\sim (\tau_\infty-\tau)^{-2(q-1/2)/q}$,
where $\tau_\infty $ is the observer proper time needed to reach
$|z| = \infty$, which is finite.
{}From the above expression we can see that the tidal forces experienced by
the freely-falling
test particles are infinitely large as the hypersurfaces  $|z| = \infty$ are
approaching. This means that these surfaces are real space-time
singularities for $q > 1/2$, instead of horizons [5]. Since in
the present case all the scalar invariants are finite, we conclude that these
singularities are intermediate (or non-scalar) singularities [14, 15].
On the other hand,  the surface energy density of the wall
per surface element is given by [5, 6]
$$
\sigma = \int^{\infty}_{-\infty} {\rho(z) e^{-\Omega/2} dz} = \sqrt{\pi} k(2 +
q)
 \frac{\Gamma(1 - q/2)}{\Gamma(3/2 - q/2)},
$$
where $\Gamma(x)$ is the standard Gamma function. From this expression we
can see that $\sigma$ is a monotonically increasing function of $q$. Note that
the constant $k$ has the meaning of the energy scale. So, we have that
{\em when the surface energy density of the wall per surface element is greater
than} $\sigma_{c}\; (\equiv \sigma |_{q = \frac{1}{2}}),$ {\em the
space-time will be closed at a
finite proper distance by space-time singularities.}
As King [16] suggested, the non-scalar singularities might not be
stable against perturbations and give rise to scalar ones. The following
analysis is in favor to King's conjecture, and will show that
 they are indeed turned into scalar
ones when null fluids are present.
 In this vein, following [17] we make the substitution
$$
\{ \Omega, h\} \rightarrow \{ \Omega + a(u) + b(v), h\},
$$
in the metric
coefficients of Eq.(1), where $a(u)$ and $b(v)$ are arbitrary functions
of their indicated arguments, with $u \equiv (t + z)/\sqrt{2},$ and
$v \equiv (t - z)/\sqrt{2}.$ Then, corresponding to the new solution,
the EMT is given by
\begin{equation}
T^{\mu}_{\nu} = \rho_{1} l^{\mu} l_{\nu} + \rho_{2} n^{\mu} n_{\nu} +
\tilde{\rho} (\delta^{\mu}_{\nu} + \tilde{\xi}^{\mu} \tilde{\xi}_{\nu})
+ \tilde{\nu} \tilde{\xi}^{\mu} \tilde{\xi}_{\nu},
\end{equation}
where
$$
\rho_{1} = -\sqrt{2} k (1 - tanh(pz)) a'(u),\;\;\;
\rho_{2} = -\sqrt{2} k (1 + tanh(pz)) b'(v),$$
$$
\tilde{\rho} = e^{a(u) + b(v)}\rho, \;\;\;\;\;
\tilde\nu =  e^{a(u) + b(v)}\nu, \eqno{\rm (5)}$$
$\tilde{\xi}_{\mu} = e^{-(a+b)/2}\xi_{\mu}$, and $l_{\mu}$ and $n_{\mu}$
are two null vectors defined, respectively, by $l_{\mu} \equiv
\partial u/\partial x^{\mu} = (\delta^{t}_{\mu} + \delta^{z}_{\mu})/
\sqrt{2}, \; n_{\mu} \equiv \partial v/\partial x^{\mu} = (\delta^{t}_{\mu}
- \delta^{z}_{\mu})/\sqrt{2}.$  The function
$\rho_{1}$ represents the energy density of the null fluid moving along the
$v = Const.$ hypersurfaces, and $\rho_{2}$ represents the energy density
of the null fluid moving along the $u = Const.$ hypersurfaces.
%
To have $\rho_{1, 2}$ positive, in the following we shall assume that
$a'(u), b'(v) < 0$. Combining this assumption with Eq.(5) we can see that,
because of the backreaction of
the null fluids, the energy density and pressure of the wall become
time-dependent, and are always decreasing as the time develops.
On the other hand, Eq.(5) also shows that $\rho_{1}$
is vanishing exponentially as one leaves from the center of the wall
to the positive $z$ direction, while $\rho_{2}$ is vanishing as one
leaves from the center to the negative $z$ direction. Thus,
the new solution represents a domain wall emitting massless particles.

Corresponding to the new solution, the Kretschmann scalar can be written
as
\setcounter{equation}{6}
\begin{equation}
{\cal{R}} \equiv R^{\mu \nu \lambda \delta} R_{\mu \nu \lambda \delta}
= e^{2(a+b)}\{ {\cal{R}}_{0} + 4e^{2\Omega} (\Phi^{(0)}_{00} \rho_{1}
+ \Phi^{(0)}_{22} \rho_{2}) + 2e^{2\Omega}\rho_{1} \rho_{2} \},
\end{equation}
where ${\cal{R}}_{0}$ is the Kretschmann scalar of the
background $\{\Omega, h\}$, and now is given by ${\cal{R}}_{0} =
12k^{4}(1+q^{2})q^{-2}(cosh(pz))^{4(q-1)}.\;
\Phi^{(0)}_{00}$ and $\Phi^{(0)}_{22}$ are the components of the traceless
Ricci tensor [18] and now given by $\Phi^{(0)}_{00} = \Phi^{(0)}_{22} =
 k^{2}(1-q)/
(2q \;cosh^{2}(pz)). \; \rho_{1, 2}$ and $\Omega$ are given, respectively,
by Eqs.(5) and (2).
The first term at the right-hand side of Eq.(7) represents the backreaction
of the null fluids to the background. The second represents the interaction
of the null fluids with the matter components $\Phi^{(0)}_{00}$ and
$\Phi^{(0)}_{22}$, which now is proportional to $cosh^{2(2q-1)}(pz).$ Thus,
for $q > 1/2$ this term will become unbounded as $|z| \rightarrow
\infty$. That is, because of the interaction of the null fluids with
the background, the intermediate singularities originally appearing at
$|z| = \infty$ are now turned into scalar ones. The last term
represents the interaction between the two null fluids, which now is also
proportional to $cosh^{2(2q-1)}(pz).$ Thus, the interaction between these
two fluids also turns those intermediate singularities into scalar
ones.

Note that at the level of the Higgs scalar field, the potential corresponding
to the new solution becomes $\tilde{V}= e^{a+b}V$.  At the field
level the new solution is obtained by formally introducing a nontrivial
dependence of a
coupling ``constant" on the coordinates $u$ and $v$, while the scalar field
is kept unchanged. When the functions $a(u)\approx 0$ and $b(v)\approx 0$
 the
coupling ``constant" is effectively unity.
Note that the
above conclusions regarding to the singular behavior of the space-time at the
space-like infinity $|z| = \infty$ hold for any $a(u)$ and $b(v)$.

On the other hand, considering the covariant derivatives of the Riemann
tensor in the freely-falling frame, we find
\begin{eqnarray}
R_{(i)(j)(k)(l);(k_{1}) ... (k_{n})} & &\rightarrow
Exp\{[2(2q - 1) + 2nq]p|z|\}  \nonumber\\
& &\rightarrow (\tau_\infty -\tau)^{-[n +2(q-1/2)/q]},\;\;\;\;\;
as \;\;\; |z| \rightarrow \infty,\nonumber
\end{eqnarray}
where indices inside parentheses denote tetrad components.
Clearly, for any given $q$ the derivatives upto certain order will become
singular on the hypersurfaces $|z| = \infty$. In particular, for $q >
\frac{1}{3}$ the first-order derivatives will become unbounded. This
indicates the existence of mild singularities on these surfaces even for
the solution with $0 < q \leq \frac{1}{2}$. For example,
for $q > \frac{1}{3}$  we have that the  ``difference" of  tidal forces
become infinitely large but the integral on the surface  is still finite.
Therefore,
one cannot exclude the possibility of an extension for $ 0 < q \leq
\frac{1}{2}$. The above argument can be further justified by the following
consideration. As we shall show below, the Riemann tensor for the extended
solution is $C^{r}$ across $|z| = \infty$ with  $r > 0$. According to the
classifications given in [14], these hypersurfaces are  $C^{r}$ regular
surfaces.

Following [19] (see also [10]), we first make the coordinate transformation
\begin{equation}
u = \alpha^{-1} e^{-k(t+|z|)}, \;\;\; v = -\alpha^{-1} e^{k(t-|z|)},
\end{equation}
where $\alpha^{2} = 2k^{2}4^{-q}$. Then, in terms of $u$ and $v$ the solution
of Eqs.(1) and (2) reads
\begin{equation}
ds^2 = [1 + (-\alpha^{2}uv)^{1/q}]^{-2q}\left\{ 2dudv - 2k^{2}v^{2}( dx^2 +
dy^{2} ) \right\},
\end{equation}
where $0 < q \leq 1/2$. From Eq.(8) we can see that the
coordinate transformations are restricted to the regions $uv < 0$. To extend
the solution into the whole $(u, v)-$plane, one just simply forgets
the way how to get Eq.(9) and lets $u$ and $v$ be any values.
Regarding to such an extension,
there are two different points of view. The first is due to Cveti\v{c}
and co-workers [10], namely, considering it as two coordinate
transformations, each of which is independently performed in the regions
$z \leq 0$ and $z \geq 0$. The resulting space-time of the wall is the
gluing of these two extended spaces along the wall. The second is to consider
Eq.(8) as one, and take Eq.(9) as the complete extension of the space-time
of the wall. Physically, the latter is equivalent to identify the two extended
spaces of the former at the same values of the coordinates. In the following,
we shall adopt the second point of view. From Eq.(9) we conclude
 that in order
to extend the solution to all the values of $u$ and $v$, we must distinguish
different cases depending on the solutions of  the algebraic equations
$(-1)^{1/q}=-1,1,i$. We shall consider
$$
a)\; q = \frac{2n + 1}{2m + 1}, \;\;\;\;
b)\; q = \frac{2n + 1}{2m},\;\;\;\;
c)\; the \; rest, \;\;\;(n, m = 0, 1, 2, ...).
$$
We shall study the last case first.

{\em Case c)}. The metric coefficients of Eq.(9) in this case become
complex in the regions $uv > 0$, which indicates that Eq.(9)
can not be considered as the proper extension beyond the surfaces $uv = 0$,
and other possibilities must be
considered. One way is to set the conformal factor in Eq.(9) as
$[1 + (\alpha^{2}uv)^{1/q}]^{-2q}$ in the regions $uv > 0$. One can show that
such an extension is $C^{1}$ in the sense of [13] and the hypersurfaces
$u = 0$ and $v = 0$ are free of any kind of matter. In the
regions $uv > 0$, introducing the coordinates $t$ and $z$ via the relations
[cf. Eq.(8)]
$$
u = \alpha^{-1} e^{-k(t+|z|)}, \;\;\; v = \alpha^{-1} e^{k(t-|z|)},
$$
we find that the metric in these regions can be written as
$$
ds^{2} = cosh^{-2q}(pz)\left\{ dz^{2} - dt^{2} -
e^{2kt}(dx^{2} + dy^{2})\right\}, \;\;\; ( uv > 0),
$$
which is the continuation of the metric (1) across the hypersurfaces
$|z| = \infty$, and clearly shows that the coordinate $t$ becomes
space-like  and  $z$ time-like. Therefore, the hypersurfaces
$|z| = \infty$ are acting as Rindler horizons [20].
It should be noted that the extension in this case is not an analytic
extension, but it is maximal, in the sense that the extended space-time is
geodesically complete [13]. Other
possibilities exist,  for example, in the regions $uv > 0$, one can replace
the conformal factor in  Eq.(9) by $1$. This extension is
$C^{1}$ across $|z| = \infty$ and not analytic. The space-time
in these regions are flat and the scalar field becomes constant.

{\em Case b)}. The metric  coefficients of Eq.(9) in this case
are well defined for all the values of $u$ and $v$. It can be shown that
when $\frac{1}{q}$ is an integer, the extension is the maximal analytic
extension,
and the extended space-time is geodesically complete. When
$\frac{1}{q}$ is not an integer, the extension is only a maximal
extension but not analytic. The space-time in the extended regions
[cf. Fig.1] is asymptotically flat.

{\em Case a)}. This is the most interesting case, as after the extension it
yields a black hole space-time
structure, which is quite similar to that of the Schwarzschild space. The
extension in this case is the maximal analytic extension when
$\frac{1}{q}$ is an integer, and only a maximal extension otherwise.
To show that the space-time indeed has
a black hole structure, let us first note that
the metric coefficients become singular on the hypersurfaces
$uv = \alpha^{-2}$ in this case. On the surface  $uv = \alpha^{-2}$
 the Kretschmann scalar,
$$
{\cal{R}} \equiv R^{\mu \nu \lambda \delta} R_{\mu \nu \lambda \delta}  =
\frac{48\alpha^{4}(1+q^{2})}{q^2} \frac { (\alpha^{2}uv)^{2(1-q)/q}}{[1 -
(\alpha^{2}uv)^{1/q}]^{4(1-q)}},
$$
becomes unbounded. In other words, a space-time singularity appears in the
extended regions, II and II\'{} [cf. Fig.1].
{}From Fig.1 we can see that this singularity is spacelike.
Note that in the space-time essentially we have two walls, each of which
is located on one of
the two branches of the hyperbola $uv = - \alpha^{-2}$.
These walls are causally disconnected one from the other and behave
like the Rindler particles
[20]. The horizons at $|z| = \infty$ (or equivalently $uv = 0$) are event
horizons.
Thus, it is concluded that the solution given by Eq.(9)
in this case represents a black hole but with plane symmetry. The plane is
defined by the three Killing vectors $\partial_x,
\partial_y$, and $x\partial_y - y\partial_x$.

However, in [7] (see also [8]) Ipser and Sikivie found that all the plane
domain walls with zero-thickness are actually bubbles. The following
considerations show explicitly that this is also the case for a thick domain
wall.
Let us first note that the metric inside
the braces of Eq.(9) is flat. As a matter of fact, by performing the
following coordinate transformations
$$
T = \{ (u + v) + k^{2}v(x^{2} + y^{2})\}/\sqrt{2},\;\;
Z = \{ (u - v) + k^{2}v(x^{2} + y^{2})\}/\sqrt{2},\;\;
$$
\begin{equation}
X = -\sqrt{2}kvx, \;\;\;\;\;\;\;\;
Y = -\sqrt{2}kvy,
\end{equation}
one can bring this part to the standard Minkowski form. If we further introduce
the spherical coordinates $\{R, \theta, \varphi\}$, which are related to
the coordinates $\{T, Z, X, Y\}$ in the usual way, we find that Eq.(9)
takes the form
\begin{equation}
ds^2 = \left\{1 + \left[\frac{\alpha^{2}}{2}(R^{2} - T^{2})\right]^{1/q}
\right\}^{-2q}\{ dT^{2} - dR^{2} -R^{2}(d\theta^{2} +
sin^{2}\theta d\varphi^{2})\},
\end{equation}
where $R^{2} = X^{2} + Y^{2} + Z^{2}$. From the above expression we can
see that the space-time of the wall has spherical symmetry in the Minkowski
coordinates.  On the other hand, from Eq.(10) we have
$$
R^{2} - T^{2} = 2\alpha^{-2} e^{-2k|z|}.
$$
Thus, the wall (the center of which is at $z = 0$) essentially is
an accelerated bubble with
the constant acceleration given by $\alpha/\sqrt{2}$ in the Minkowski
coordinates. It starts to collapse at the moment $T = - \infty$
until the moment $T = 0$, where the radius of
it is $R_{min.} = \sqrt{2}/\alpha$. Since the acceleration
is outward, the wall will start to expand afterwards. Note that the physical
radius of the wall is given by $r_{ph} = 2^{-q}R = 2^{-q}(2\alpha^{-2} +
T^{2})^{1/2}$.

In summary, we have studied the local and global properties of the
Goetz thick plane domain wall solution. It has been found that for
$1/2 < q < 1$, intermediate
singularities appear on the hypersurfaces $|z| =
\infty$. When  null fluids are present, these
singularities become scalar ones. The solution with $0 < q \leq 1/2$
has been extended beyond the horizons $|z| = \infty$. The
extended space-time for some choice of the free parameter $q$ has a black
hole structure.
Moreover, we have
shown  explicitly that the space-time of the walls in the Minkowski
coordinates
has spherical symmetry, instead of the plane symmetry as  originally presented.
The geometry of the wall is a spherically symmetric closed surface, which is
collapsing initially. However, because of its constant outward acceleration,
when it collapses to its minimum radius, which is always greater than zero,
the wall starts to expand.

\vspace {1.cm}

\noindent {\bf{Acknowledgments}}

Part of the work was done when one of the authors (A.W.) was visiting the
Department of Applied Mathematics, UNICAMP. He thanks the hospitality of
the Department.
This work was partially supported by a grant from FAPESP and a grant from
CNPq.

\newpage

\noindent {\bf\Large{ Figure Caption}}

Fig.1 The projection of the space-time onto the $uv-$plane. The
hypersurfaces $|z| = +\infty$ ($uv = 0$) are event
horizons. The center of the walls is at $z = 0$ ($uv = -\alpha^{-2}$).
For Case a) where $q = (2n+1)/(2m+1)$, the space-time is singular on
$uv = \alpha^{-2}$, and consequently regions II and II\'{} represent
two catastrophic regions. Regions II, II\'{}, III and III\'{} are the four
extended regions that are missing in the (t, z) coordinates, where III $\equiv
\{ uv < -\alpha^{-2}, u > 0 \},$ and III\'{} $\equiv
\{ uv < -\alpha^{-2}, v > 0 \}.$

\end{document}